\documentclass[twocolumn,tighten]{aastex63}

\usepackage{amsmath}
\usepackage{xspace}
\usepackage{multirow}
\usepackage{fancyapj}
\usepackage[modulo]{lineno}

\newcommand*\patchAmsMathEnvironmentForLineno[1]{%
  \expandafter\let\csname old#1\expandafter\endcsname\csname #1\endcsname
  \expandafter\let\csname oldend#1\expandafter\endcsname\csname end#1\endcsname
  \renewenvironment{#1}%
     {\linenomath\csname old#1\endcsname}%
     {\csname oldend#1\endcsname\endlinenomath}}%
\newcommand*\patchBothAmsMathEnvironmentsForLineno[1]{%
  \patchAmsMathEnvironmentForLineno{#1}%
  \patchAmsMathEnvironmentForLineno{#1*}}%
\AtBeginDocument{%
\patchBothAmsMathEnvironmentsForLineno{equation}%
\patchBothAmsMathEnvironmentsForLineno{align}%
\patchBothAmsMathEnvironmentsForLineno{flalign}%
\patchBothAmsMathEnvironmentsForLineno{alignat}%
\patchBothAmsMathEnvironmentsForLineno{gather}%
\patchBothAmsMathEnvironmentsForLineno{multline}%
}

  {\list{}{\leftmargin=0.1in\rightmargin=0.1in}\item[]}%
  {\endlist}

\expandafter\def\csname editcolor1\endcsname{magenta}

\newcommand{\br}[1]{\ensuremath{\left[ #1 \right]} }
\newcommand{\rbr}[1]{\ensuremath{\left( #1 \right)} }

\newcommand{\E}[1]{\ensuremath{\times 10^{#1}} }

\newcommand{\msol}{\ensuremath{M_{\odot}}\xspace}
\newcommand{\ah}{\ensuremath{^{\rm h}}}
\newcommand{\am}{\ensuremath{^{\rm m}}}
\newcommand{\as}{\ensuremath{^{\rm s}}}
\newcommand{\s}{{\rm\,s}\xspace}
\newcommand{\cts}{ct\,\per{s}\xspace}
\newcommand{\hz}{{\rm\,Hz}\xspace} 
\newcommand{\kev}{{\rm\,keV}\xspace} 
\newcommand{\km}{{\rm\,km}\xspace} 
\newcommand{\per}[1]{{\rm\,#1\ensuremath{^{-1}}}\xspace}
\newcommand{\persq}[1]{\,#1\ensuremath{^{-2}}\xspace}

\newcommand{\tasc}{\ensuremath{T_{\rm asc}}\xspace}
\newcommand{\rchi}{\ensuremath{\chi^2_{\rm r}}\xspace}
\newcommand{\rmag}{\ensuremath{r_{\rm m}}\xspace}
\newcommand{\rcor}{\ensuremath{r_{\rm c}}\xspace}
\newcommand{\mdot}{\ensuremath{\dot M}\xspace}
\newcommand{\kpc}{{\rm\,kpc}\xspace} 

\newcommand{\rxte}{\textit{RXTE}\xspace}
\newcommand{\xmm}{\textit{XMM-Newton}\xspace}

\newcommand{\chandra}{\textit{Chandra}\xspace}
\newcommand{\nustar}{\textit{NuSTAR}\xspace}
\newcommand{\nicer}{\textit{NICER}\xspace}

\newcommand{\src}{SAX~J1808\xspace}

\begin{document}

\title{Timing the pulsations of the accreting millisecond pulsar SAX J1808.4--3658
during its 2019 outburst}

\author{Peter Bult}
\affiliation{Astrophysics Science Division, 
  NASA's Goddard Space Flight Center, Greenbelt, MD 20771, USA}

\author{Deepto Chakrabarty}
\affil{MIT Kavli Institute for Astrophysics and Space Research, 
  Massachusetts Institute of Technology, Cambridge, MA 02139, USA}

\author{Zaven Arzoumanian} 
\affiliation{Astrophysics Science Division, 
  NASA's Goddard Space Flight Center, Greenbelt, MD 20771, USA}

\author{Keith C. Gendreau} 
\affiliation{Astrophysics Science Division, 
  NASA's Goddard Space Flight Center, Greenbelt, MD 20771, USA}

\author{Sebastien Guillot} 
\affil{CNRS, IRAP, 9 avenue du Colonel Roche, BP
  44346, F-31028 Toulouse Cedex 4, France} 
\affil{Universit\'e de Toulouse, CNES, UPS-OMP, F-31028 Toulouse, France}

\author{Christian Malacaria}
\altaffiliation{NASA Postdoctoral Fellow}
\affiliation{NASA Marshall Space Flight Center, 
  NSSTC, 320 Sparkman Drive, Huntsville, AL 35805, USA}
\affiliation{Universities Space Research Association, \
  NSSTC, 320 Sparkman Drive, Huntsville, AL 35805, USA}

\author{Paul. S. Ray}
\affiliation{Space Science Division, Naval Research Laboratory,
  Washington, DC 20375-5352, USA}

\author{Tod E. Strohmayer} 
\affil{Astrophysics Science Division and Joint Space-Science Institute,
  NASA's Goddard Space Flight Center, Greenbelt, MD 20771, USA}

\begin{abstract}
    In this paper we present a coherent timing analysis of the 401\,Hz pulsations
    of the accreting millisecond X-ray pulsar SAX J1808.4--3658 during its 2019
    outburst. Using observations collected with the \textit{Neutron Star
    Interior Composition Explorer} (\nicer), we establish the pulsar spin
    frequency and orbital phase during its latest epoch. 
    We find that the 2019 outburst shows a pronounced evolution in pulse phase
    over the course of the outburst. These phase shifts are found
    to correlate with the source flux, and are interpreted in terms of 
    hot-spot drift on the stellar surface, driven by changes in the mass
    accretion rate. Additionally, we find that the long-term evolution of the
    pulsar spin frequency shows evidence for a modulation at the Earth's
    orbital period, enabling pulsar timing based astrometry of this
    accreting millisecond pulsar. 
\end{abstract}


\section{Introduction}
  \label{sec:intro}
  The low-mass X-ray binary SAX~J1808.4--3658 (hereafter \src) is the
  canonical example of an accreting millisecond X-ray pulsar (AMXP). First
  discovered as an X-ray source in 1996 with the \textit{BeppoSAX} satellite
  \citep{Zand1998}, it was not until 1998 that the detection of 401\hz
  pulsations with the \textit{Rossi X-ray Timing Explorer} (\rxte) established
  this source as the first known AMXP \citep{Wijnands1998a}. Over the two
  decades since this discovery, the sample size of known AMXPs has grown
  substantially \citep{Patruno2012b, Campana2018}.  Despite this growth,
  however, \src remains the cornerstone of its class.  This status is in part
  due to its relatively close proximity, at an estimated distance of $3.5\kpc$
  \citep{Galloway2006}, and in part due to its short, semi-regular outburst
  recurrence times of $2-4$ years. Together, these qualities have enabled 
  extensive investigations of its pulsations
  \citep{Poutanen2003, Hartman2008,Burderi2009,Patruno2017,Sanna2017c}, its aperiodic
  variability \citep{Wijnands2001, Patruno2009c, Bult2015a, Bult2015b}, its
  spectral characteristics \citep{Gierlinski2002, Cackett2009,
  Papitto2009, DiSalvo2019}, and its thermonuclear X-ray bursts
  \citep{Zand2001, Chakrabarty2003, Galloway2006, Bhattacharyya2007, Zand2013}.

  The characteristic property of an AMXP is its coherently pulsed emission.
  These oscillations have nearly sinusoidal profiles that are attributed
  to a localized emission region, a hot-spot, on the neutron star surface.
  Such a hot-spot arises due to the dynamically relevant stellar magnetosphere,
  which channels in-falling plasma, so that the accretion column impacts the
  stellar surface in confined regions near the magnetic poles. These pulsations
  reveal a host of information about the neutron star, with the precise
  waveform encoding information on the stellar compactness
  \citep{Poutanen2003}, and their phase evolution with time revealing the
  spin evolution of the binary and star itself \citep{Hartman2008}. 
  
  For \src specifically, coherent timing analyses of the pulsations revealed
  that the X-ray outbursts do not show a measurable accretion-powered spin
  change \citep{Hartman2008,Hartman2009,Patruno2012a,Patruno2017,Sanna2017c},
  while between outbursts the source spins down at a steady rate of
  $\approx1.5\E{-15}\hz\per{s}$. Additionally, these analyses revealed
  the $2.01$\,hr binary orbit \citep{Chakrabarty1998},
  and established a peculiar long-term variation of the binary
  period. The first decade of monitoring showed that the binary period
  was expanding at an anomalously fast rate of $\approx3.5\E{-12}\s\per{s}$
  \citep{DiSalvo2008, Hartman2008}, which was explained as being due to
  highly non-conservative mass-transfer \citep{DiSalvo2008, Burderi2009},
  or, alternatively, due to spin-orbit angular momentum exchange in the companion 
  star \citep{Hartman2008, Patruno2012a}, akin to short-term orbital
  variability observed in millisecond ``black widow" radio pulsars \citep{Arzoumanian1994,
  Doroshenko2001}.  Later outbursts, in 2011 and 2015, showed that the rate of
  orbital expansion is itself evolving rapidly. These measurements have been
  analysed in great detail \citep{Patruno2017, Sanna2017c}, but for a lack of
  data points, no single conclusion could yet be reached.

  On 2019 July 30, \citet{ATelRussell19} reported a brightening in the
  optical i'-band from the direction of \src, which is often a precursor
  to a full X-ray outburst \citep{Russell2019}.
  Although this predictor had not previously been tested for \src
  specifically, various X-ray, UV, and radio observatories were
  triggered to monitor \src over the following days. On 2019 August 6
  a further increase in the optical i'-band was observed, coincident
  with a flux increase in the X-ray band \citep{ATelGoodwin19} and UV
  \citep{ATelParikh19d}, suggesting the source was indeed starting a new
  outburst. On 2019 August 7 the X-ray flux started increasing
  rapidly, and the detection of 401\,Hz pulsations \citep{ATelBult19c} with the
  \textit{Neutron Star Interior Composition Explorer} (\nicer;
  \citealt{Gendreau2017}) confirmed that \src was indeed active again. 

  Following confirmation of a new outburst, we executed an extensive,
  high cadence monitoring campaign with \nicer. In this paper we report on a
  timing analysis of the coherent pulsations for these data.  

\section{Observations}
\label{sec:observations}
  We observed \src with \nicer between 2019 July 30 and
  2019 August 31 for a total unfiltered exposure of $355.4$\,ks.
  These data are available under program/target numbers $205026$ and $258401$.  
  At the time of the final observation included in the present analysis,
  the source had completed the high luminosity phase of its outburst cycle,
  but remained visible in its characteristic prolonged flaring tail
  \citep{Wijnands2001,Patruno2009c,Patruno2016a, ATelBult19d}.  

  We processed all data using \textsc{nicerdas} version 6a, which is
  distributed as part of \textsc{heasoft} v6.26. We applied standard
  screening criteria, limiting our analysis to time intervals with a
  pointing offset $<54\arcsec$, a bright Earth limb angle $>30\arcdeg$, 
  a dark Earth limb angle $>15\arcdeg$, and collected outside the South
  Atlantic Anomaly. 
  The standard processing also applies a background screening filter, 
  which rejects all time intervals where the rate of saturating particle 
  events (overshoots) is greater than $1$\,\cts\per{detector},
  or greater than 1.52 $\times$ \textsc{cor\_sax}$^{-0.633}$, where
  \textsc{cor\_sax}\footnote{%
      The \textsc{cor\_sax} parameter is based on a model for the cut-off
      rigidity that was originally developed for the \textit{BeppoSAX}
      satellite. It has no relation to the source \src.
  } gives the cut-off rigidity of the Earth's magnetic field, in units
  of GeV\per{c}.  For the \src observations, we find that this approach
  is often too conservative, as statistical fluctuations in the
  overshoot rate are leading to many unnecessary $1-10\s$ gaps
  in the light curve. We therefore applied a smoothing average filter
  to the overshoot light curve using a 5 second window. Additionally,
  we relaxed the cut-off threshold for the overshoots to
  $1.5$\,\cts\per{detector} and increased the scaling factor of the
  \textsc{cor\_sax} expression from 1.52 to 2.0.  A visual comparison of
  light curves based on standard and relaxed filtering showed that no
  spurious signals were introduced in this way.
  Using these criteria, we retained {257.7}\,ks of good time
  exposure.  These good time data were corrected to the Solar System
  barycenter using the \textsc{barycorr} tool, where we used the
  optical coordinates of \citet{Hartman2008} and the JPL DE405 planetary
  ephemeris \citep{Standish1998}. All times reported in this work are
  therefore expressed in units of Barycentric Dynamical Time (TDB).
  Finally, we estimated the background contributions to our data from
  \nicer observations of the \rxte blank-field regions
  \citep{Jahoda2006}.

  Two thermonuclear X-ray bursts were observed with \nicer, the first on
  August 9 and the second on August 21. We exclude both events from the
  analysis presented here. An initial analysis of the second bursts is 
  presented in \citet{Bult2019b}. An analysis of the first burst, and a
  separate, detailed review of the spectroscopic and stochastic variability
  will be presented elsewhere.

\section{Coherent Timing}
\label{sec:results}

  We performed a coherent timing analysis of the 401\hz 
  pulsations of \src. For each continuous good-time interval 
  we corrected the photon arrival times for the Doppler delays of the
  binary motion based on the preliminary ephemeris of
  \citet{ATelBult19c}. We then folded the data on the pulsar period to
  construct a pulse profile. This profile was decomposed into its
  Fourier components by fitting a constant plus $k$ harmonic sinusoids.
  Each sinusoid thus has a fixed frequency $\nu_k = k \nu_p$, where
  $\nu_p$ is the pulsar spin frequency. Additionally, for each sinusoid
  we determined the phase, $\varphi_k$, and amplitude, $A_k$, from which
  we derive a fiducial pulse arrival time and a fractional sinusoidal
  amplitude, $r$, as
  \begin{equation}
    r = \frac{A_k}{N_\gamma - B},
  \end{equation}
  where $N_\gamma$ gives the number of photons in the pulse profile, and
  $B$ gives the estimated number of photons contributed by the background
  emission. A harmonic was deemed significant if its measured amplitude
  exceeded the $99\%$ confidence amplitude threshold of the noise
  distribution. When a harmonic was not significantly detected, we
  instead computed an upper limit on the amplitude as the minimum signal
  that would have produced a measured amplitude in excess of the noise
  threshold $95\%$ of the time. 
  
\begin{figure}[t]
  \centering
  \includegraphics[width=\linewidth]{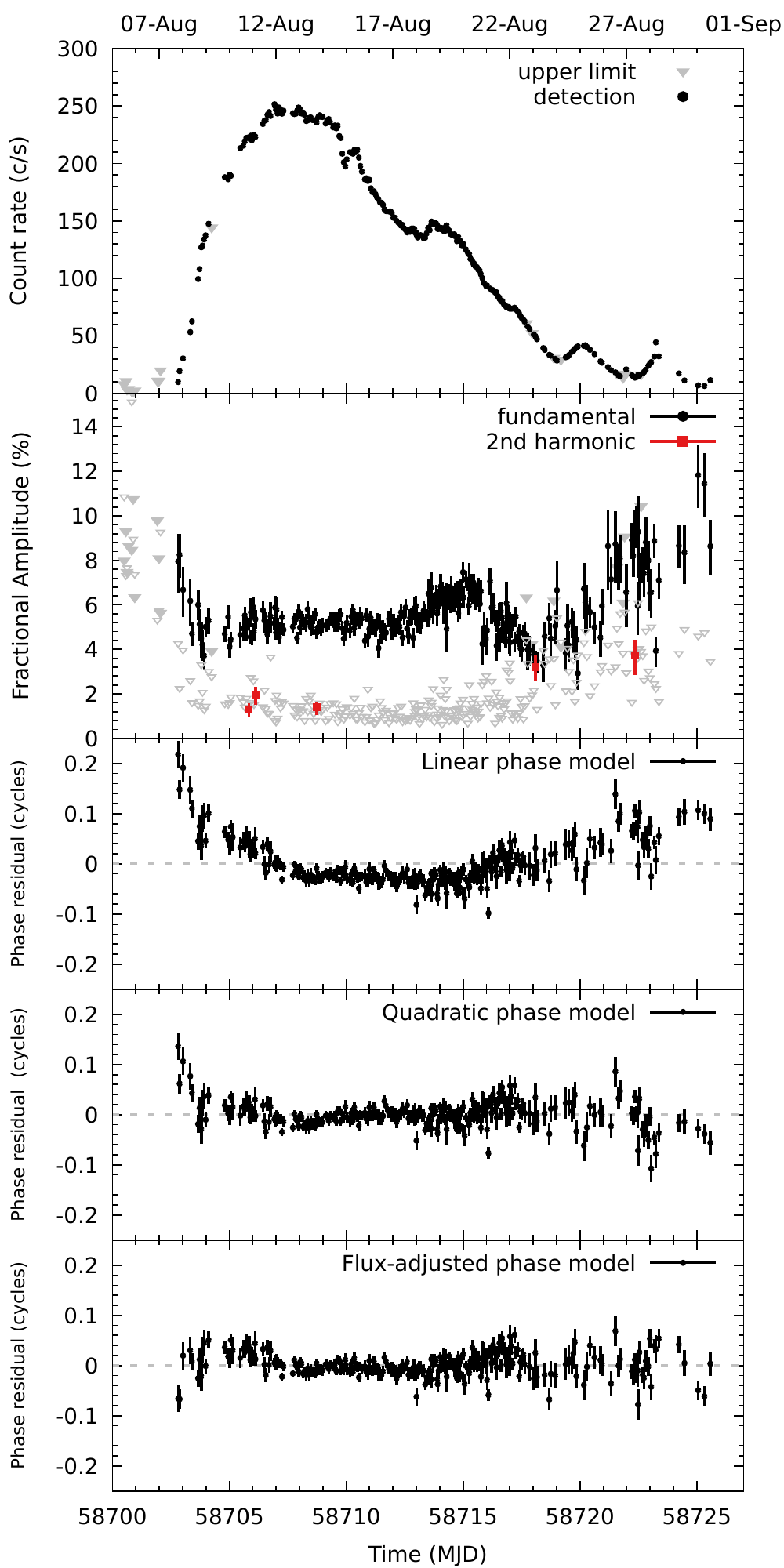}
  \caption{%
      Temporal evolution of the 2019 outburst of \src. Top panel: the
      $0.5-10$\kev light curve. Second panel: pulse fractional sinusoidal amplitudes,
      with solid triangles giving the upper limits for non-detections of the
      fundamental component, and open triangles the upper limits for the second
      harmonic.  Third through fifth panels: pulse phase residuals relative to the
      indicated phase model (see also Table \ref{tab:ephemeris}).
  }
  \label{fig:pulse evolution}
\end{figure}

  Pulsations at the fundamental frequency are detected at high
  significance throughout the outburst (Figure \ref{fig:pulse
  evolution}, 1st and 2nd panel). For a number of segments,
  however, the integration time was too short to reach a detection.
  Given the small number of cases, we did not attempt to combine these
  segments with neighboring intervals, and show an upper limit on
  the amplitude instead. The second harmonic was rarely detected, mainly
  because its amplitude is small below 3\kev
  \citep{Patruno2009a}, where \nicer is most sensitive. Indeed, by
  accumulating more exposure in each pulse profile, we could
  detect the second harmonic significantly in more segments. However,
  the required integration times were long ($>5$\,ks), greatly reducing  
  the number of measured pulse arrival times. Hence, we focus our
  analysis on the arrival times of the fundamental pulsation. 
  
  The series of pulse arrival times was modeled as a constant
  spin frequency and a circular Keplerian orbit using \textsc{tempo2}
  \citep{Hobbs2006}. Hence, our initial model consisted of four
  parameters: the binary orbital period, $P_b$, the projected semi-major
  axis, $a_x \sin i$, the time of passage through the ascending node, $\tasc$,
  and the spin frequency, $\nu_p$.
  After an initial fit, we repeated the epoch-folding method using the 
  refined ephemeris, and repeated the entire procedure until the 
  solution converged. 
  
  At a reduced $\chi^2$ (\rchi) of 7.2 for 259 degrees of freedom (dof),
  we find that our initial model provided a poor description to the 
  set of arrival times, leaving an approximately parabolic trend in the
  phase residuals (Figure \ref{fig:pulse evolution}, 3rd panel).  This
  trend suggests that the spin frequency is changing over the course
  of the outburst.  Indeed, including a spin frequency derivative in our
  timing model improves the fit statistic to a \rchi of 2.27 for 258
  dof. Considering the fit residuals, however, we find large phase
  residuals near the start and end of our data set (Figure
  \ref{fig:pulse evolution}, 4th panel). This suggests that additional,
  higher spin frequency derivatives, for which there is little physical
  motivation, are needed to describe the data

  An important consideration in the coherent timing analysis of
  accreting millisecond pulsars, however, is that the pulse profiles can
  change shape, leading to both gradual and abrupt shifts in the
  observed phase \citep[see][for a review]{Patruno2012b}. Indeed, in
  \src the pulse waveform is known to be sensitive to the source flux
  \citep{Patruno2009b}.  
  We therefore generalized our timing model by
  expressing the pulsar phase as
  \begin{equation}
    \varphi(t,F) = \varphi_0 + \nu_p t + \varphi_{\rm orb}(t) +
    \varphi_{\rm bias}(F_{\rm bol}),
  \end{equation}
  where $\varphi_0$ gives a reference phase, $\nu_p t$ gives the linear
  term due to the constant spin frequency, $\varphi_{\rm orb}(t)$ gives the phase
  correction associated with the binary orbit, and
  $\varphi_{\rm bias}(F_{\rm bol})$
  gives a phase-bias correction term as a function of
  bolometric flux, $F_{\rm bol}$. In order to specify a bias function, we note
  that numerical simulations indicate that the azimuthal position of the
  hot-spot is proportional to the magnetospheric radius \citep{Kulkarni2013}.
  Additionally, the magnetospheric radius is predicted to scale with the mass
  accretion rate as ${\dot M}^{-1/5}$ \citep{dAngelo2010, Kulkarni2013}.  Since
  $\dot M$ should be proportional to the bolometric flux, we define a simple
  power-law model
  \begin{equation}
    \label{eq:bias}
      \varphi_{\rm bias}(F_{\rm bol}) = b F_{\rm bol}^{\Gamma},
  \end{equation}
  with $\Gamma$ fixed at $-1/5$, and $b$ a free scaling factor, which
  we will refer to as the scale of the bias. Finally, for simplicity, we use
  count-rate as an approximate substitute for the flux. Applying this
  generalized model to our arrival times, we obtain a \rchi of 2.08 for 258
  dof. While this fit statistic is still not formally acceptable, it performs
  better than the quadratic phase model. Furthermore, we note that
  phase residuals for this model (Figure \ref{fig:pulse evolution}, 5th panel)
  are no longer structured, such that the poor statistic can be attributed to
  residual timing noise \citep{Patruno2012b}. 

  Finally, we note that our data quality is sufficient that the
  Keplerian parameters describing the binary orbit are effectively
  decoupled from the parameters describing the spin frequency
  evolution. Indeed, in all three models discussed above, the Keplerian
  parameters agree within their respective uncertainties. 
  Hence, in Table \ref{tab:ephemeris} we list the best-fit timing
  solution for the orbit separately from spin frequency parameters
  obtained from the three phase models.

\begin{table}[t]
    \newcommand{\mc}[1]{\multicolumn3l{#1}}
    \caption{%
        Timing solution for the 2019 outburst of \src. 
        \label{tab:ephemeris}
    }
    \begin{center}
      \begin{tabular}{p{2.5cm} L l}
        \tableline
        Parameter & \phm{-}{\rm Value} & {Uncertainty} \\
        \tableline
        Epoch (MJD)           &\phm{-}58715               & - \\
        $a_x \sin i$ (lt-ms)  &\phm{-}62.8091             & 7.4\E{-3} \\
        $P_{b}$ (s)           &\phm{-}7249.1552           & 3.0\E{-3} \\
        $T_{\rm asc}$ (MJD)   &\phm{-}58715.0221031       & 2.0\E{-6} \\
        $\epsilon$            & <3\E{-4}            & \\ 
        \tableline
        \mc{~~\it Linear phase model} \\
        \tableline
        $\nu$ (Hz)            &\phm{-}400.975209741       & 1.2\E{-8} \\
        $\chi^2$/dof          &\phm{-}1854.79/259         & ~ \\  
        \tableline
        \mc{~~\it Quadratic phase model} \\
        \tableline
        $\nu$ (Hz)            & \phm{-}400.975210055  & 2.8\E{-8} \\
        $\dot\nu$ (Hz\per{s}) & -3.02\E{-13}          & 1.3\E{-14} ~    \\
        $\chi^2$/dof          & \phm{-}585.30/258     & ~ \\
        \tableline
        \mc{~~\it Flux-adjusted phase model} \\
        \tableline
        $\nu$ (Hz)            &\phm{-}400.975209833   & 1.0\E{-8} \\
        $b$                   & -0.87                 & 0.03 \\
        $\Gamma$              & -1 / 5                & fixed \\
        $\chi^2$/dof          &\phm{-}535.95 / 258    & ~   \\
        \tableline
    \end{tabular}
    \end{center}
    \tablecomments{%
        Uncertainties give the $1\sigma$ statistical errors
        and the upper limit is quoted at the 95\% c.l. The $\epsilon$ 
        parameter gives the orbital eccentricity.
    }
\end{table}

\section{Pulse energy dependence}
\begin{figure*}[t]
  \centering
  \includegraphics[width=0.47\linewidth]{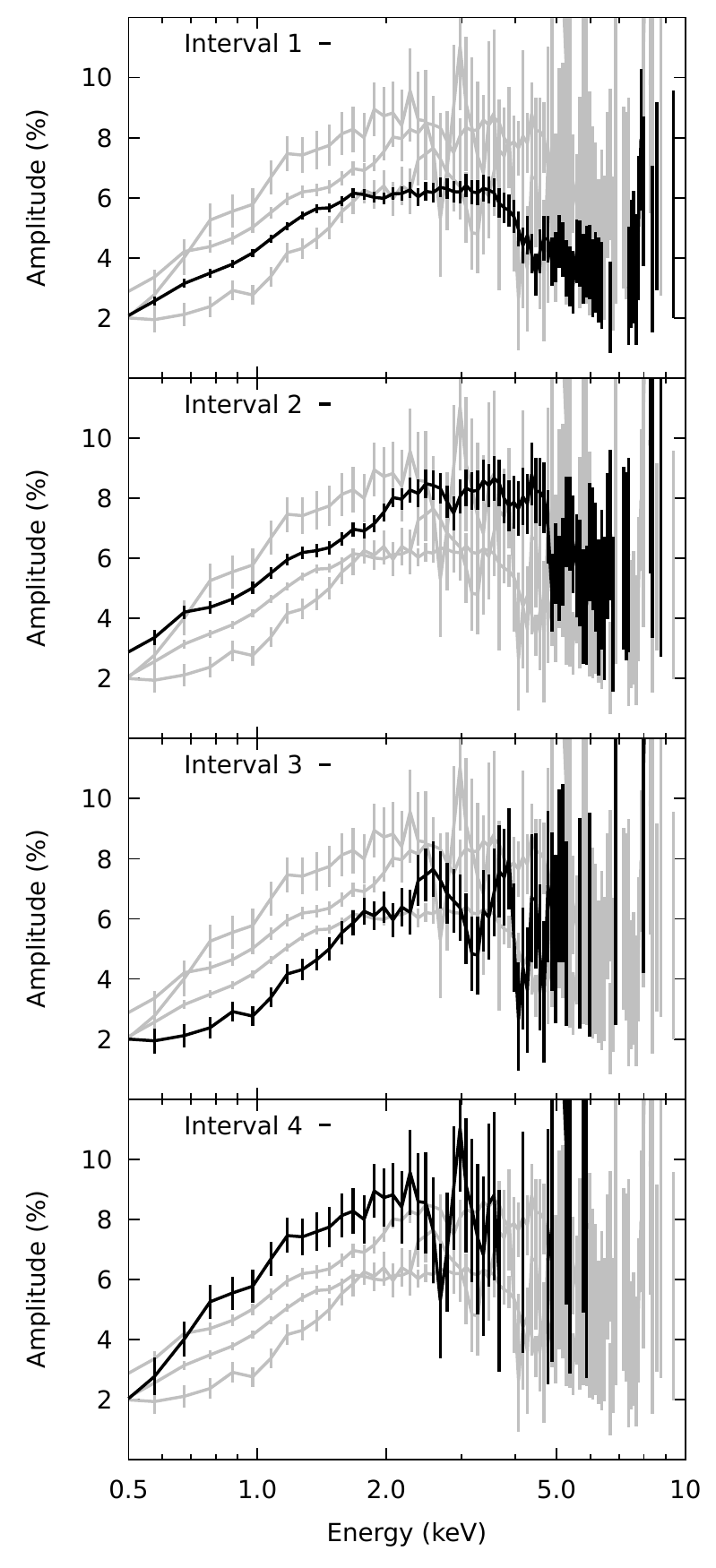}
  \includegraphics[width=0.47\linewidth]{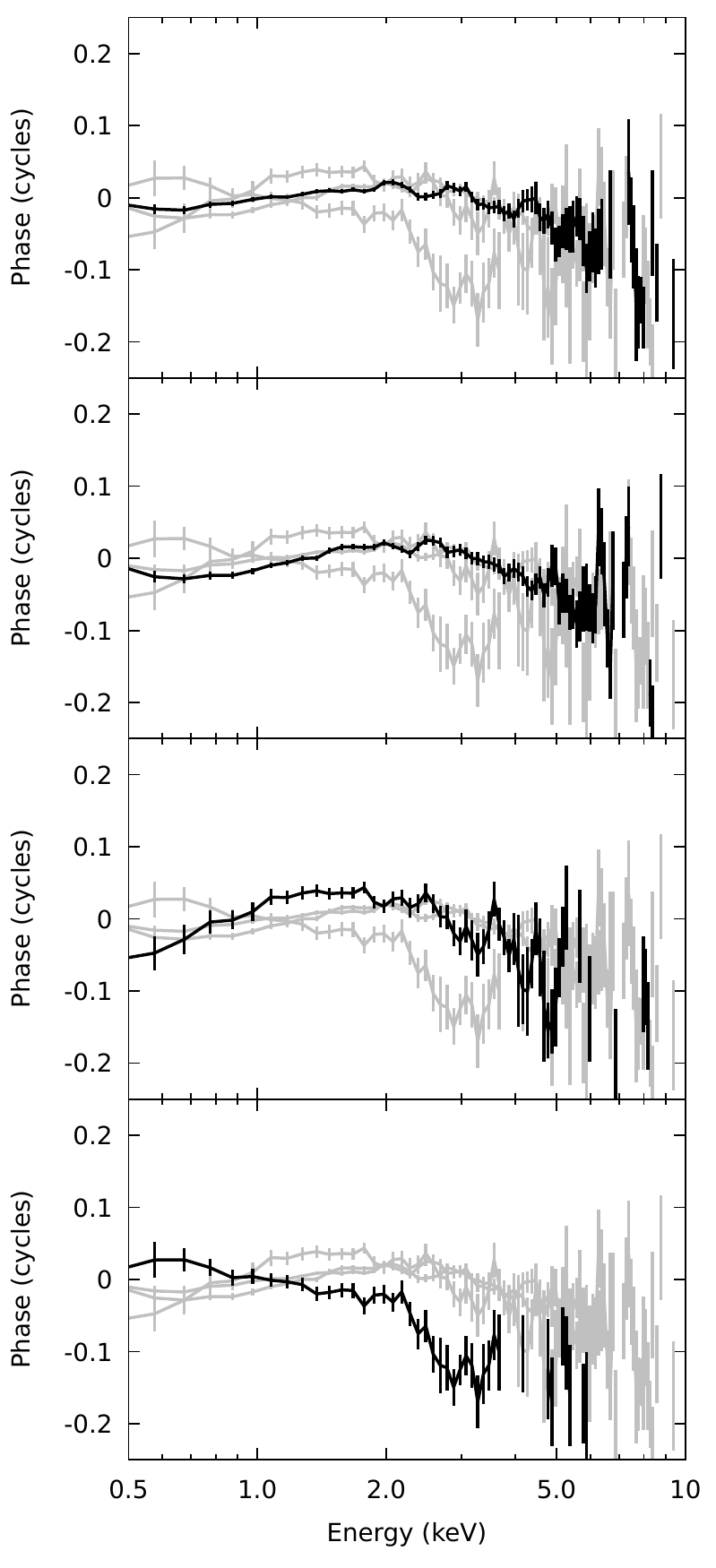}
  \caption{%
    Energy resolved pulse fractional amplitude (left) and phase
    (right). Each panel shows a spectrum for the indicated time interval
    in black, with the other three intervals plotted in grey for context. 
  }
  \label{fig:pulse spectrum}
\end{figure*}

  To investigate the photon-energy dependence of the pulsations, we
  applied a sliding window method in energy space. That is, we selected
  all data in a $0.25\kev$ wide window and measured the resulting pulse
  profile properties. We then repeated this analysis by moving the
  window from $0.5\kev$ to $10\kev$ using steps of $0.1\kev$.
  This method was applied to four representative time intervals.
  The first interval covers August 10 through 17, which is when the
  outburst peaked and the broad-band fractional pulse amplitude remained
  constant at $5\%$. 
  The second interval covers August 18 through 20, which is where the
  source flux showed a re-brightening with respect to its long-term
  decay, and, likewise, the broad-band fractional pulse amplitude temporarily
  increased to $7\%$.
  The third interval covers August 21 through 25, where the outburst
  continued its decay and the fractional pulse amplitude varied between
  $4\%$ and $7\%$.
  Finally, the fourth group covers August 26 through 31 where the source began
  to transition to its flaring state, and the broad-band fractional pulse
  amplitude increased above $7\%$, indicating an associated transition
  in the pulse waveform.
  The resulting pulse amplitude and phase spectra are
  shown in Figure \ref{fig:pulse spectrum}.

  Consistent with previous results \citep{Patruno2009a,Sanna2017c}, we
  find that the pulse amplitude increases with energy up to
  $\approx2\kev$, and then flattens off (top panel). More interesting,
  however, is to note that the overall pulse energy spectrum shows
  significant evolution with time. As the source briefly re-brightens 
  during its decay (second panel), the amplitude spectrum hardens, but
  the phase remains mostly the same. However, as the source intensity drops
  further, the phase residuals below $2\kev$ begin to pivot, as
  the fractional amplitudes below $2\kev$ show large variations (third
  and bottom panels).

\section{Long term evolution}

\subsection{Spin frequency}
\label{sec:evolution}
\begin{figure}[t]
  \centering
  \includegraphics[width=\linewidth]{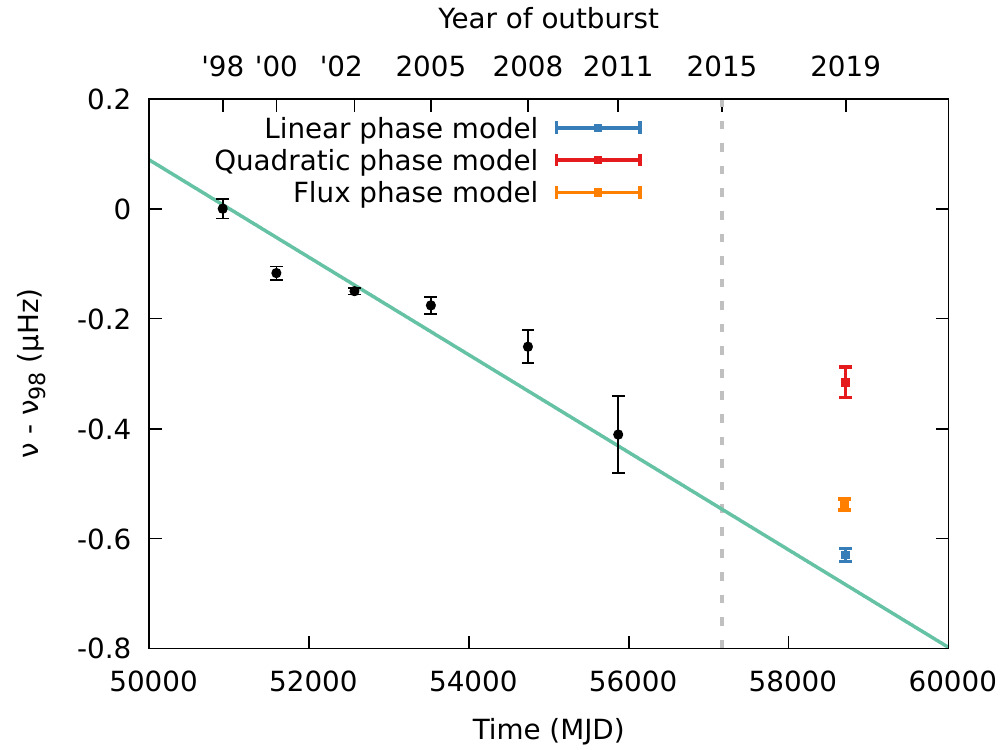}
  \caption{%
    Spin frequency evolution of \src as measured over two decades.
    Black bullets mark measurements made with \rxte. The colored squares
    give the \nicer measurements for three phase models (see text). 
    All frequencies are expressed relative to the 1998 epoch, with $\nu_{98} =
    400.975210371$\,Hz. The dashed grey line marks the reference time of the
    2015 outburst, for which an high accuracy pulse frequency
    measurement is not available (see text). The solid line indicates
    the best fit frequency evolution as determined in Section
    \ref{sec:evolution}.  
  }
  \label{fig:freq evolution}
\end{figure}

  Between 1998 and 2011, the spin frequency of \src decreased at a
  constant rate (\citealt{Patruno2012a}; for the 2015 outburst, see
  discussion). In Figure \ref{fig:freq evolution} we
  compare the three spin frequencies measured for the 2019 data with
  this established spin-down trend. The historic data for this comparison
  was adopted from \citet{Hartman2008,Hartman2009} and \citet{Patruno2012a}.
  The frequencies of the linear and flux-adjusted phase models both confirm a
  continuation of the long-term spin-down. The frequency derived from the
  quadratic phase model, however, lies well away from the trend.  To
  quantitatively model the spin frequency evolution, we adopt the frequency
  from the linear phase model, as that model is the most similar to the
  historic data (see Section \ref{sec:spin discussion}).

  While there is a clear decreasing trend in the long term evolution of
  the spin frequency, a linear function does not provide a statistically
  acceptable description of these data (\rchi of 5.1 for 5 dof). If
  we remove the long-term linear spin-down trend, and consider the
  residuals as a function of day of year, we observe an apparently
  sinusoidal modulation at the Earth's orbital period. 
  This result suggests that our source coordinates are slightly offset
  from the optical position, such that the barycenter correction leaves
  a small periodic Doppler shift in the data \citep{Manchester1972, Manchester1974,
  Lyne1990}.  
  
\begin{figure}[t]
  \centering
  \includegraphics[width=\linewidth]{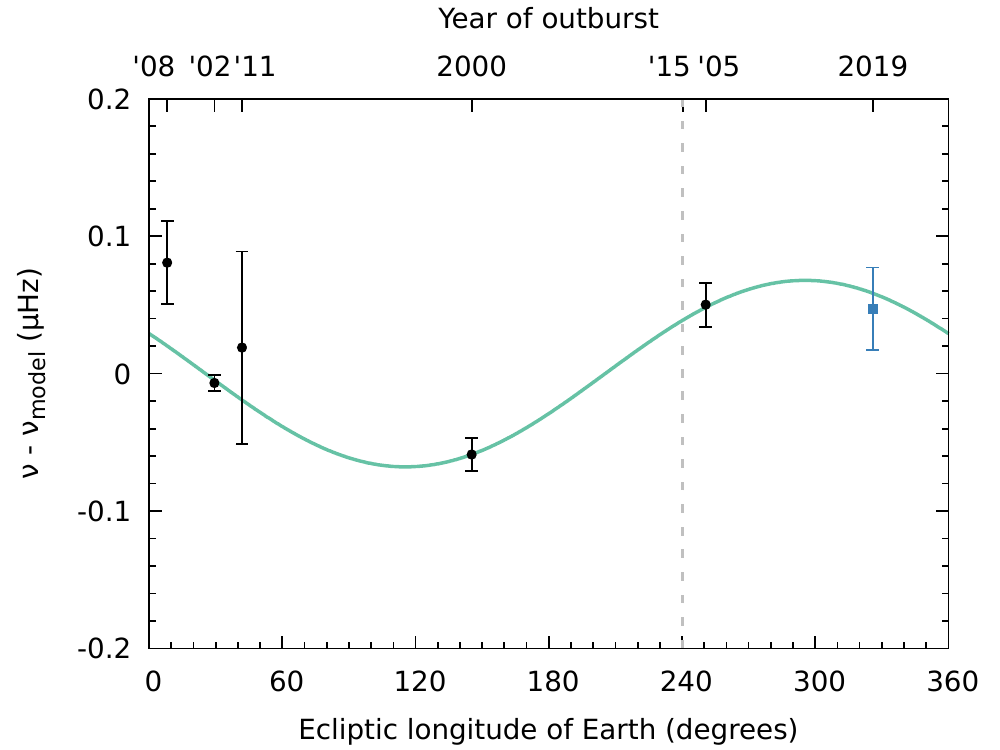}
  \caption{%
    Spin frequency measurements relative to the best-fit spin-down model
    (see also Figure \ref{fig:freq evolution}) as a function of the Earth's
    ecliptic longitude.
  }
  \label{fig:astrometry}
\end{figure}

  To account for the uncertainty in the source coordinates, we modeled
  the long-term evolution of the spin frequency as a linear trend plus a
  correction term for the residual Doppler shift,
  \begin{equation}
    \Delta\nu(t)
      = \delta\nu_{98} + \dot\nu(t-T_{98})
      + \mathcal{D}(t, \delta\lambda, \delta\beta),
  \end{equation}
  where $\Delta\nu$ gives the measured spin frequency relative to
  $\nu_{98}$, the spin frequency measured for the epoch of the 1998
  outburst, $T_{98}$. Additionally, $\delta\nu_{98}$ gives the
  correction to the spin frequency, $\dot\nu$ the 
  spin frequency derivative at $T_{98}$, and $\mathcal{D}(t, \delta\lambda,
  \delta\beta)$ the Doppler correction term, which is defined as
  \begin{align}
    \label{eq:doppler}
    \mathcal{D}(t, \delta\lambda, \delta\beta) 
    &= \nu_p \frac{a_{\oplus}}{c} \omega 
    \left[ \vphantom{\frac{1}{2}}
    \delta\lambda \cos(\beta) \cos(\omega (t-T_\Upsilon) - \lambda)  \right. \nonumber \\
    &+ \left. \delta\beta \sin(\beta) \sin(\omega (t-T_\Upsilon) - \lambda)
      \vphantom{\frac{1}{2}}
      \right],
  \end{align}
  where $a_\oplus$ is the Earth's circular orbital radius, $c$ is the speed of light,
  $\omega=2\pi/P_\oplus$ is the Earth's circular angular frequency,
  $P_\oplus$ is a sidereal year, and $T_\Upsilon= {\rm MJD~} 51079.143$ is the epoch
  of zero degrees ecliptic longitude for the Earth's orbit\footnote{
      This occurs when the Earth lies in the direction of the J2000 vernal
      equinox ($\Upsilon$) relative to the Sun, which is approximately the date
      of the northern autumnal equinox (September 23).
  }.
  Finally, $\lambda$ and $\beta$ give the
  respective source longitude and latitude in ecliptic coordinates,
  with $\delta\lambda$ and $\delta\beta$ their correction terms. 
  We further note that corrections associated with the Earth's orbital
  eccentricity are negligible, and can be safely ignored. 
  Applying this model to the data, we obtain a good fit (\rchi of 1.7
  for 3 dof) for a coordinate offset of $\delta\lambda = 0.33\arcsec \pm
  0.10\arcsec$ and $\delta\beta = -0.60\arcsec \pm 0.25\arcsec$. In
  Figure \ref{fig:astrometry} we show how this model fits the spin
  frequency residuals relative to the linear spin-down trend.

  Our timing coordinates are offset from the optical position by a
  $2.3\sigma$ deviation. This offset is sufficiently large that one might
  wonder if the source shows measurable proper motion. To test for
  proper motion, we substituted the constant coordinate offsets in
  Equation \ref{eq:doppler} with their linear expansion in time, i.e.,
  \begin{align}
    \delta\lambda(t) &= \delta\lambda_0 + \mu_\lambda ( t - T_{\rm optical}) \\
    \delta\beta(t) &= \delta\beta_0 + \mu_\beta ( t - T_{\rm optical}),
  \end{align}
  where $T_{\rm optical} = {\rm MJD~} 52703$ gives the 2001 June 13 date on which
  the optical data were collected \citep{Hartman2008}. Assuming the
  optical position was exactly correct (i.e., fixing $\delta\lambda_0$
  and $\delta\beta_0$ to zero), we obtain a very poor fit (\rchi of 7.1
  for 3 dof).  
  Letting all parameters vary yields a \rchi of 1.8 for 1 dof, with mean offset
  of $\delta\lambda_0 = 0.33\arcsec \pm 0.10\arcsec$ and $\delta\beta_0 =
  -0.37\arcsec \pm 0.40\arcsec$, and with ill-constrained derivatives of
  $\mu_\lambda = +0.013\pm 0.10 \arcsec\per{yr}$ and $\mu_\beta = -0.04
  \pm 0.03 \arcsec\per{yr}$. Hence, the long-term spin frequency measurements
  are best described using a fixed position offset.  We transform the pulsar
  timing source position to equatorial coordinates, and list them in Table
  \ref{tab:long term}, along with the associated spin frequency and its
  derivative. 
  
  Finally, we note that similar results can be obtained by adopting the spin
  frequency obtained from the flux-adjusted phase model. In this case the long
  term spin down was slightly smaller, at $\dot\nu = (-8.5 \pm
  0.5)\E{-16}\hz\per{s}$, while the measured offset in ecliptic latitude
  reduced to $\delta\beta = -0.40\arcsec \pm0.19\arcsec$ and the offset in ecliptic
  longitude was $\delta\lambda = 0.28\arcsec \pm 0.07\arcsec$.

\begin{table}
  \caption{
    Long term timing solution of \src.
    \label{tab:long term}
  }
  \begin{tabular}{l L l}
    \tableline
      Parameter & \phm{-}{\rm Value} & {Uncertainty} \\
    \tableline
      R.A.                    & \phm-18\ah 08\am 27\fs647             & 0.007\as \\
      decl.                   & -36\arcdeg 58\arcmin 43\farcs90 & 0.25\arcsec \\
      Epoch (MJD)             & \phm-50921.6                         & - \\
      $\nu$ (\hz)             & \phm-400.975210371                   & 1.4\E{-8} \\
      $\dot\nu$ (\hz\per{s})  & -1.01\E{-15}                         & 7\E{-17} \\
    \tableline
  \end{tabular}
  \tablecomments{%
      Uncertainties give the $1\sigma$ statistical errors. Both the spin frequency
      and its derivative at are defined relative to the epoch of the 1998 outburst.
  }
\end{table}

\subsection{Orbital period}
\begin{figure}[t]
  \centering
  \includegraphics[width=\linewidth]{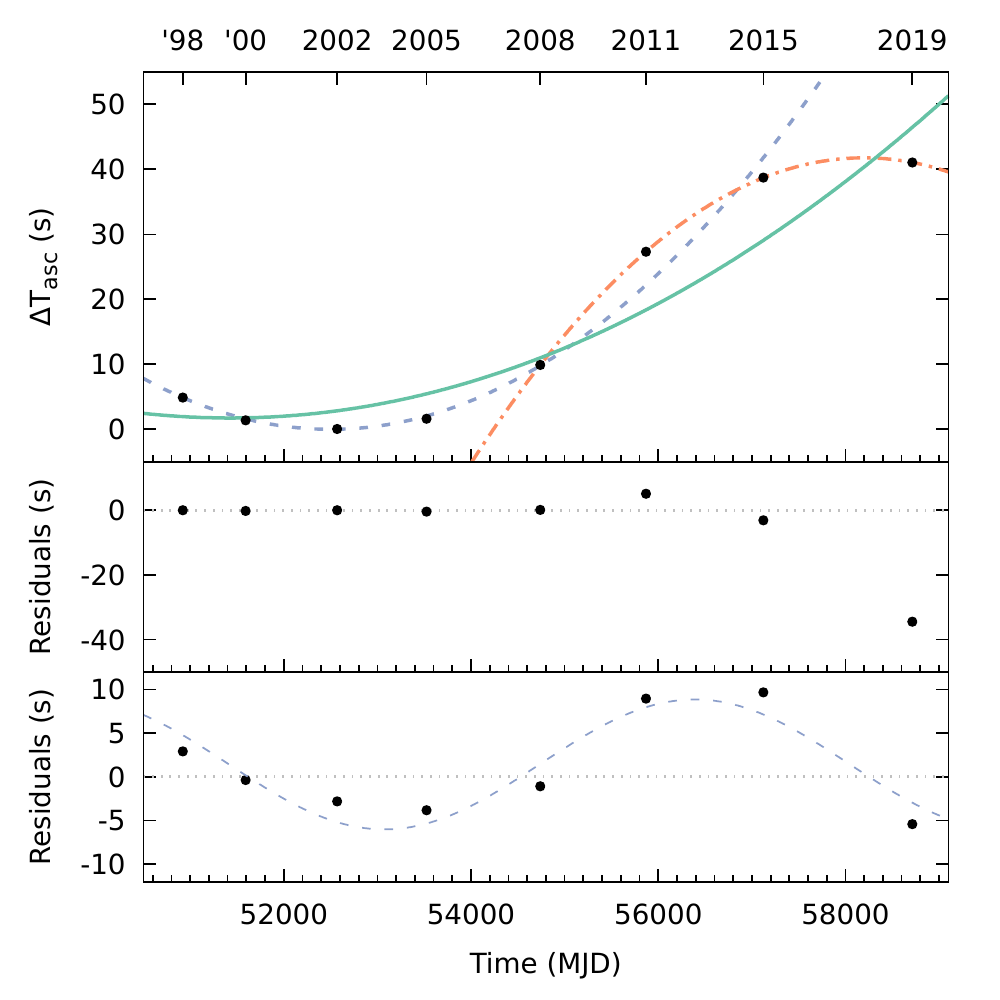}
  \caption{%
    Top: Residual offset of the \tasc measurements relative to predicted times
    based on the 2002 orbital period and reference time \citep{Hartman2009}.
    The curves show parabolic trends fit to the $1998-2008$ (dashed), $2008-2019$
    (dash-dot), and $1998-2019$ (solid) subsets of the data.
    Middle: residuals relative to the $1998-2008$ parabolic model.  Bottom:
    residuals relative to a parabolic model fit to the entire dataset. A
    sinusoid with a $18.2$\,yr period and $7\s$ amplitude was added to guide
    the eye. In all panels the uncertainties are smaller than the point size. 
  }
  \label{fig:orbit evolution}
\end{figure}

  To study the long-term evolution of the orbital period, we follow the procedure
  of \citet{Hartman2008} and calculate the residual time of passage
  through the ascending node, $\Delta\tasc$, as 
  \begin{equation}
    \Delta\tasc = T_{{\rm asc}, i} - (T_{\rm ref} + N P_b),
  \end{equation}
  where we set the reference time to the epoch of the 2002 outburst,
  $T_{\rm ref} = 52499.96$. Additionally,
  $T_{{\rm asc}, i}$ is the time of ascending node measured for the $i$th
  outburst, $N$ is the number of orbital revolutions between the $i$th outburst
  and $T_{\rm ref}$, and $P_b$ gives the orbital period reported by
  \citet{Hartman2009}.  These residual delays are shown in Figure
  \ref{fig:orbit evolution}. For context, we also show the parabolic model of
  \citet{Hartman2009}, and the residuals with respect to this model.

  The $\Delta\tasc$ measured for the 2019 outburst represents a significant
  lead with respect to the model of \citet{Hartman2009}. From Figure 
  \ref{fig:orbit evolution} we see that this offset is a continuation of 
  a recent trend, which suggests that the binary orbit has been contracting in 
  the recent decade. Indeed, when we consider only the second decade of measurements
  (2008 through 2019), then we can fit the $\Delta\tasc$ evolution with a
  parabolic function (\rchi of 0.02 for 1 dof). This fit yields an orbital
  period derivative of $\dot P_b = (-5.18 \pm 0.02)\E{-12}\s\per{s}$. 

  Considering all data, we find that the mean long-term trend of the
  $\Delta\tasc$ yields a period derivative of $\dot P_b = (1.6 \pm
  0.7)\E{-12}\s\per{s}$. The residuals around this mean trend show a large
  amplitude oscillation with a period of approximately $18.2$\,years. This
  oscillation is clearly not sinusoidal, and could also not be adequately
  described as time delays due to a third body in an eccentric orbit. Instead,
  this quasi-periodic oscillation is apparently a stochastic process that is
  intrinsic to the dynamics of the binary orbit \citep[see, e.g.,][for detailed
  modeling]{Patruno2017,Sanna2017c}. What is not clear from the available data
  is whether this quasi-periodic process has a $\approx7\s$ amplitude around a
  steadily expanding orbit, or if we are seeing a $\approx20\s$ amplitude
  modulation around a constant binary period. Additional monitoring of future
  outbursts is needed to differentiate between these two scenarios.

\section{Discussion}
\label{sec:discussion}
We have presented a coherent timing analysis of the 2019 outburst
of \src as observed with \nicer. We found that over the course of the
outburst, the pulse phase changes as a function of the count-rate. 
Considering the long-term evolution, we found there is now sufficient
data for pulsar timing astrometry. In the following we consider some
of the implications of these results.

\subsection{Spin frequency evolution}
\label{sec:spin discussion}
  We modeled the long-term evolution of the neutron star spin frequency by
  taking into account possible frequency shifts due to the position
  uncertainty. We found that this model gave a statistically acceptable
  description of the measurements, implying that we obtained an improved source
  position through pulsar timing astrometry. The timing position obtained
  through this analysis is consistent with the optical at a $2.3\sigma$ level.
  Given that the $0.15\arcsec$ uncertainty in the optical position is set
  by the systematic uncertainty in the 2MASS catalog \citep{Hartman2008},
  we investigated if the relatively large offset between the optical and pulsar
  timing positions could be explained through proper motion. However, we found
  that no significant proper motion could be measured from the pulsar timing
  data.

  An important caveat to this analysis is that the averaged spin
  frequency measured in each outburst is subject to systematic modelling
  uncertainty (as is clear from Table 1).  For consistency, we only used spin
  frequency measurements that were obtained using a linear phase model 
  \citep{Hartman2008, Hartman2009}.  While \citet{Patruno2009b} report slightly offset spin
  frequency measurements
  using a flux correction, there is an important distinction between the way we
  modeled the flux-adjustment and their work. \citet{Patruno2009b} determined the
  optimal timing solution by minimizing a linear fit to the phase-flux
  correlation, as a function of pulse frequency.  In contrast, we incorporated
  the bias term directly into the phase model through a physically motivated 
  power-law dependence.  Hence, the reference frame of our flux-adjusted phase
  model is not guaranteed to be the same as that of \citet{Patruno2009b}.
  Because a complete reanalysis of the archival \rxte data is beyond the scope
  of this work, we applied our spin-down analysis to the linear phase model.
  While this choice may have introduced some bias into our best-fit solution,
  the main implication of our analysis is that we now have a sufficiently long
  baseline of measurements for \src that astrometric effects can no longer be
  treated as uncertainties, but instead should be accounted for in the timing
  analysis.

  Finally, we note that we did not include a spin frequency measurement from
  the 2015 outburst.  While a spin frequency was determined for this outburst
  using \xmm and \nustar \citep{Sanna2017c}, as well as with \chandra
  \citep{Patruno2017}, each of these measurements suffers from substantial
  statistical and systematic uncertainty.  Specifically, the \xmm and \nustar
  data suffered from clock drifts, which limited the absolute time resolution.
  The \chandra data, on the other hand, was collected late in the outburst,
  when the source had already decayed to low flux levels. These data have
  relatively poor statistical uncertainty on the frequency measurements.
  Additionally, all these observations only sample a very limited flux range,
  and are potentially biased by the flux-phase relation.
  
\subsection{Hot-spot drift}
    We found that, over the course of the 2019 outburst, the pulse phase showed
    significant evolution relative to a constant spin frequency model. We
    attempted to account for this evolution with two approaches: a quadratic
    phase model, and a flux-adjusted phase model.

    The quadratic phase model accounts for the phase drift by adding a spin
    frequency derivative to the timing model. The physical interpretation of
    this model is that accretion torques applied during outburst are measurably
    affecting the spin frequency of the neutron star. While this model mostly
    accounts for the observed phase drift, there are significant issues with
    it. First, the residuals still show a (minor) systematic trend that is not
    consistent with timing noise. Second, the spin frequency derivative obtained
    through this model is anomalous. At $-3\E{-13}\hz\per{s}$, the derivative
    implies that the accretion torque is spinning-down the neutron star.
    Previous outbursts did not show such a spin-down. Although some early work on
    \src report spin changes on the order of $10^{-13}\hz\per{s}$ \citep{Morgan2003,
    Burderi2007}, the sign and magnitude of the derivative change depending on
    how the data are selected, based on which \citet{Hartman2008} argued that the
    observed phase changes are actually rooted in the shape changes of the
    pulse waveform.  
    Third, we expect that the accretion torque acts in the same manner for
    every outburst, so by rescaling the apparent spin-down measured in 2019
    based on the outburst recurrence cycle, we obtain a long-time averaged
    accretion-powered spin down of $-6\E{-15}\hz\per{s}$.  This rate is
    significantly higher than the long-term spin down that is actually
    observed.  Furthermore, the mean 2019 spin frequency obtained from this
    model is actually larger than the mean spin frequency measured during the
    2011 outburst, further exacerbating this inconsistency.  Given these
    contradictions, we conclude that the quadratic phase model is not physical. 

    In our second approach, we accounted for the phase shifts by adding a
    flux-dependent adjustment to our timing model. The physical interpretation for this phase
    model is that the location of the hot-spot on the stellar surface is not
    fixed. Instead, the hot-spot drifts as a function of mass accretion rate,
    causing apparent phase shifts that are not related to the stellar rotation
    \citep{Romanova2003,Lamb2009,Patruno2009b}.
    To describe the hot-spot drift, we adopted a power-law term in our phase
    model (Equation \ref{eq:bias}). This flux-adjusted model performs
    significantly better than the quadratic model discussed above. The spin
    frequency obtained through this model is larger than that of the linear
    phase model (by 90\,nHz, or $6\sigma$), but still broadly consistent with
    the long term spin-down trend.   

\begin{figure}[t]
  \centering
  \includegraphics[width=\linewidth]{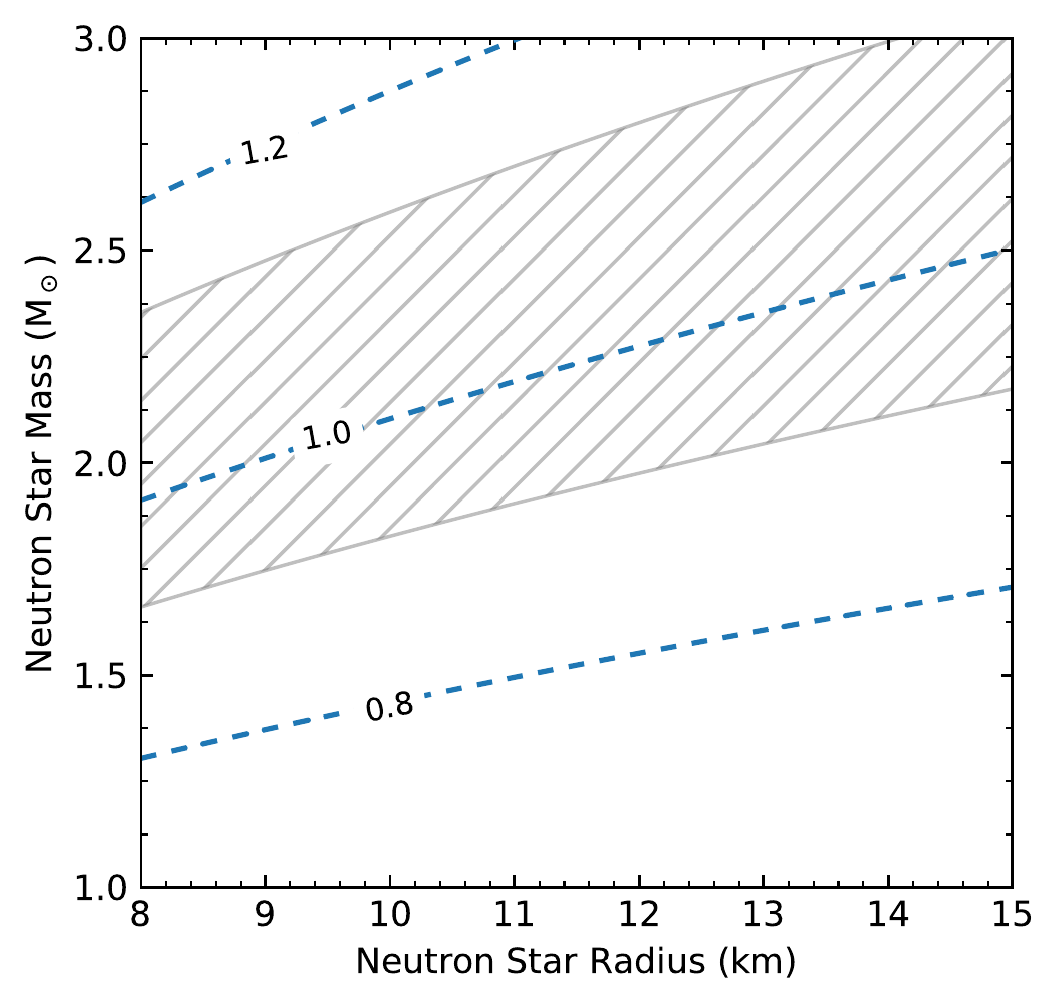}
  \caption{%
      Neutron star mass radius constraints from the bias scale of the
      flux-adjusted phase model. The dashed lines show iso-contours for the
      magnetic dipole model parameter $\mu$ in units of $10^{26}{\rm\,G\,cm}^{3}$, 
      assuming a distance of $3.5\kpc$. The hatched region illustrates the
      distance uncertainty, for $\mu$ fixed at $1.0\E{26}{\rm\,G\,cm}^{3}$.
  }
  \label{fig:mass radius mu}
\end{figure}

    Our choice for a power-law dependence, and its $-1/5$ index, was guided by
    numerical simulations of accretion onto magnetic stars \citep{Kulkarni2013}.
    These simulations indicate that the azimuthal position of the hot-spot
    relative to the magnetic pole scales linearly with the magnetospheric
    radius as 
    \begin{equation}
        \label{eq:phase prediction}
        \phi_0 \propto -135\arcdeg \frac{\rmag}{\rcor},
    \end{equation}
    where \rmag is the magnetospheric radius and \rcor the co-rotation radius.
    The latter is defined as the radius at which the Keplerian orbital frequency
    matches the spin frequency, 
    \begin{equation}
        \rcor = \rbr{\frac{GM}{\Omega^2}}^{1/3},
    \end{equation}
    with $G$ the gravitational constant, $M$ the neutron star mass,
    and $\Omega=2\pi\nu_p$ the angular spin frequency of the neutron star.
    The former is defined as the inner truncation radius of the accretion disk,
    which follows from balancing the magnetospheric stresses against the
    material stresses in the accretion disk.  The classical definition is
    \citep{Pringle1972, Lamb1973}
    \begin{equation*}
        \rmag = \gamma_B^{2/7} \rbr{ \frac{\mu^4}{G M \mdot^2} }^{1/7},   
    \end{equation*}
    where $\mu$ is the magnetic dipole moment and $\gamma_B$ captures
    the uncertain physics determining the extent of the boundary layer where
    the disk couples to the magnetosphere \citep{Psaltis1999b}.  Numerical
    simulations, however, indicate a weaker dependence on mass accretion rate
    \citep{Kulkarni2013},
    \begin{equation}
        \label{eq:kr13}
        \rmag = 1.06 R \rbr{ \frac{\mu^2}{G M \mdot R^7} }^{1/5},
    \end{equation}
    with $R$ the neutron star radius.
     The accretion rate can be estimated from the bolometric flux through
     the relation $4 \pi d^2 F_{\rm bol} = G M \mdot / R$. In Section \ref{sec:results}
     we further assumed that the bolometric flux, $F_{\rm bol}$, is proportional to 
     the observed count-rate. To calibrate this relation, we fit
     each continuous good time interval using an absorbed multi-temperature
     disk blackbody and a Compontized power-law \citep[see also,][]{Papitto2009,
     DiSalvo2019, Bult2019b}. For each spectrum, we then estimated the
     bolometric flux by extrapolating the model between 0.01\kev and 100\kev.
     The resulting fluxes were indeed nearly proportional to the $0.5-10$\kev
     count-rate, $\Lambda$, with
     \begin{equation}
         \label{eq:flux scale}
         F_b = \kappa \Lambda,
     \end{equation}
     where we measured $\kappa=4.14\E{-12}$\,erg\persq{cm}\per{ct}.
     Rewriting these expressions, we find, for the azimuthal position of the
     hot-spot,
     \begin{equation}
         \phi_0 \propto -135\arcdeg \beta \Lambda^{-1/5},
     \end{equation}
     where,  
     \begin{align}
         \label{eq:beta}
         \beta &= 2.92 
         \br{\frac{M}{1.4\msol}}^{-7/30}
         \br{\frac{R}{10\km}}^{1/10}
         \nonumber \\ &\times 
         \br{\frac{\mu}{1.4\E{26}{\rm\,G\,cm}^3}}^{2/5}
         \br{\frac{d}{3.5\kpc}}^{-2/5},
     \end{align}
    Thus, from theory, we expect that the scale of our count-rate
    dependent bias should be $b = -1.09$, which is offset by about $20\%$ from
    the $b = -0.87$ that we actually measured using our flux-adjusted
    phase model (Table \ref{tab:ephemeris}). This offset can be
    accounted by varying the neutron star parameters within their
    allowed ranges \citep{Galloway2006,Hartman2008,Goodwin2019}, but only if
    $\mu<1.2\E{26}{\rm\,G\,cm}^{3}$ (see Figure \ref{fig:mass radius mu}).

    The magnetic moment of \src has been constrained to
    $(0.7-1.4)\E{26}{\rm\,G\,cm}^3$ \citep{Hartman2008,Patruno2012a}. The upper
    limit follows from relating the long-term spin down trend to 
    magnetic dipole radiation. The lower limit is obtained by requiring that,
    at the highest luminosities, the magnetic field be strong enough to truncate
    the disk at the neutron star radius \citep{Psaltis1999b}, although we
    note this value changes depending on specific assumptions for the source distance and 
    the neutron star parameters \citep[see, e.g.,][for an overview]{Mukherjee2015}.
    In this context, the upper limit on $\mu$ that we obtained through the
    bias scale may indicate that the magnetic dipole radiation is not the only
    source of torque applied to the neutron star. Instead, an additional
    torque may be present in the form of the propeller mechanism
    \citep{Illarionov1975, dAngelo2010}, or gravitational wave emission due to
    a neutron star mass quadrupole \citep{Wagoner1984, Bildsten1998b,
    Andersson2001, Melatos2005, Mahmoodifar2013}. 

\begin{figure}[t]
  \centering
  \includegraphics[width=\linewidth]{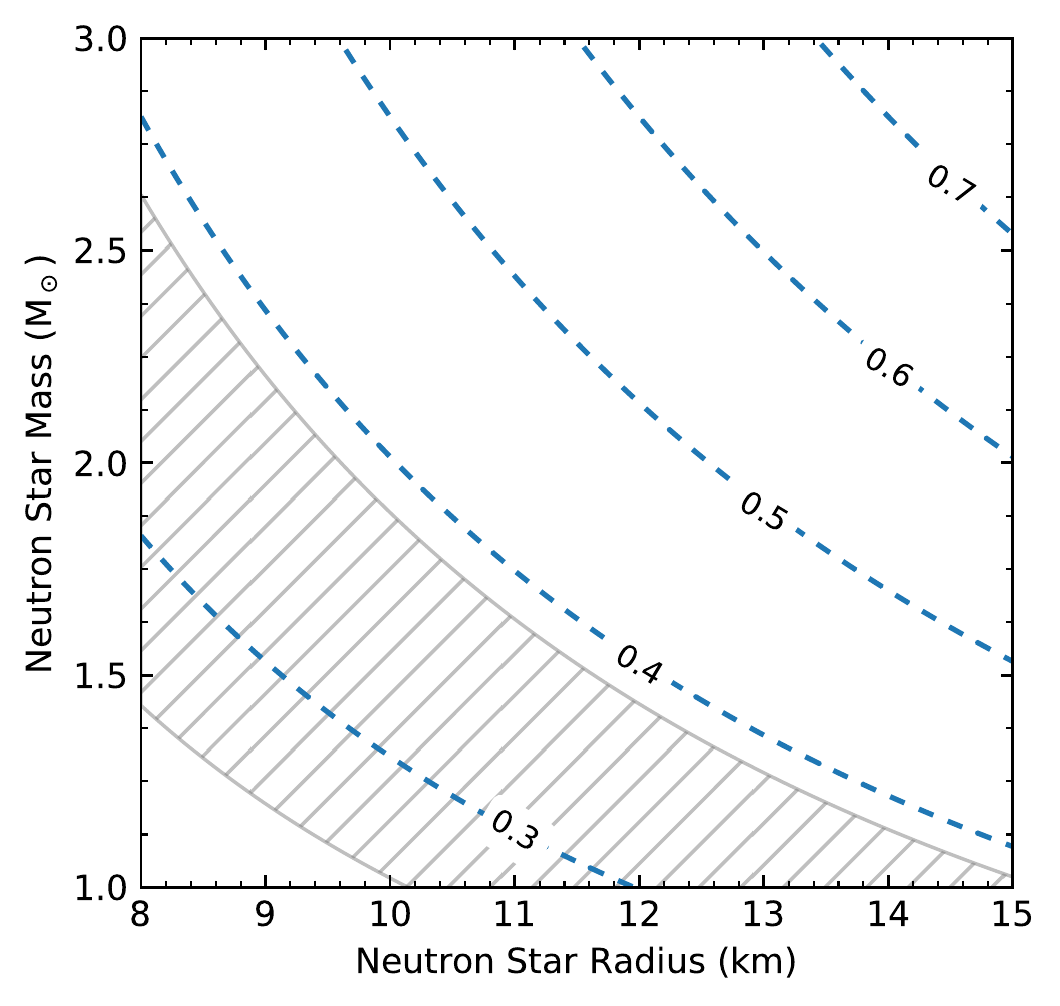}
  \caption{%
      Neutron star mass radius constraints from the bias scale of the
      flux-adjusted phase model. The dashed lines show iso-contours for the
      magnetic diffusivity parameter $\eta$, assuming a distance of $3.5\kpc$. The
      hatched region illustrates the distance uncertainty, for $\eta$ fixed at 0.3.
      The magnetic moment has been fixed to $\mu=1.4\E{26}{\rm\,G\,cm}^3$.
  }
  \label{fig:mass radius eta}
\end{figure}

    Alternatively, the tension between the magnetic moment from the spin-down
    trend and that from the bias scale could be attributed to uncertainties
    related to the disk-magnetosphere boundary region. By determining the point
    at which the magnetic field is strong enough to force the disk into co-rotation,
    \citet{dAngelo2010} derive the magnetospheric radius as
    \begin{equation}
        \label{eq:ds10}
        \rmag = \rbr{\frac{\eta \mu^2}{4 \Omega \mdot}}^{1/5},
    \end{equation}
    where $\eta$ parameterizes the poorly understood magnetic diffusivity of
    the accretion disk.  We point out that this relation almost exactly
    matches Equation \ref{eq:kr13} when $\eta=1$ and the neutron star mass and
    radius are fixed to the values used in the simulations of
    \citet{Kulkarni2013}. The different dependence on $R$ is due to the assumption
    of what the orbital velocity at the inner edge of the disk should be. Using
    equation \ref{eq:ds10} we can derive an alternative expression for $\beta$
    as
    \begin{align}
         \label{eq:beta}
         \beta &= 2.92 
         \times \eta^{1/5}
         \br{\frac{M}{1.4\msol}}^{-2/15}
         \br{\frac{R}{10\km}}^{-1/5}
         \nonumber \\ &\times 
         \br{\frac{\mu}{1.4\E{26}{\rm\,G\,cm}^3}}^{2/5}
         \br{\frac{d}{3.5\kpc}}^{-2/5}.
    \end{align}
    Assuming magnetic dipole radiation is the dominant source of spin-down
    torque (i.e., setting $\mu=1.4\E{26}{\rm\,G\,cm}^3$), we can constrain $\eta$
    to $\approx0.3-0.4$ (see Figure \ref{fig:mass radius eta}).
    
    While these constraints are subject to the various uncertainties in the
    numerical modelling, the fact that we can successfully describe the
    systematic timing noise using a physical model provides a strong argument
    in favor of this interpretation.  Additionally, that we can obtain
    reasonable values for the properties of the neutron star bodes well for
    future efforts in detailed waveform modelling of accreting millisecond
    pulsars. In this analysis we only consider the phase residuals, but
    additional information is encoded in the precise shape of the waveform
    \citep{Poutanen2003, Ibragimov2009, Morsink2011}, and in its energy
    dependence. Hence, comprehensive time-resolved waveform modelling of \src
    may not only reveal the neutron star mass and radius, but also constrain
    the detailed physics of the disk-magnetosphere boundary region.

\facilities{ADS, HEASARC, NICER}
\software{heasoft (v6.26), nicerdas (v6a), tempo2 \citep{Hobbs2006}}
    
\nolinenumbers
\acknowledgments
This work was supported by NASA through the \nicer mission and the
Astrophysics Explorers Program, and made use of data and software 
provided by the High Energy Astrophysics Science Archive Research Center 
(HEASARC).
P.B. was supported by an NPP fellowship at NASA Goddard Space Flight Center.

\bibliographystyle{fancyapj}

\end{document}